\documentclass[11pt]{article}
\usepackage{aas_macros,amsmath,amssymb,comment,cite,esint,graphicx,mathtools}
\usepackage[margin=.8in,letterpaper]{geometry}
\usepackage[colorlinks=true]{hyperref}
\usepackage[affil-it]{authblk}
\usepackage{subcaption}
\usepackage[utf8]{inputenc}
\usepackage{mathrsfs}
\usepackage{appendix}
\usepackage{amssymb}
\usepackage{float}
\usepackage{color}
\usepackage{cite}
\usepackage{hyperref}
\usepackage{longtable}
\hypersetup{pageanchor=false}
\usepackage{indentfirst}
\usepackage{url}
\usepackage{float}
\usepackage{caption}
\usepackage[numbers,square,comma,sort&compress,merge]{natbib}
\usepackage{esint}
\usepackage{overpic}
\usepackage{graphicx}
\usepackage{epsf,amsmath,bbold,amsfonts,stmaryrd}
\usepackage{textcomp}
\usepackage{ulem}
\usepackage{tikz}
\usepackage{multirow}

\numberwithin{equation}{section}
\let\originalleft\left
\let\originalright\right
\renewcommand{\left}{\mathopen{}\mathclose\bgroup\originalleft}
\renewcommand{\right}{\aftergroup\egroup\originalright}
\mathcode`\*="8000
{\catcode`\*=\active\gdef*{\mathclose{}\,\mathopen{}}}

\title{How to describe the Sweet-Parker model in general relativity}

\author{
Ye Shen,$^{1}$\thanks{E-mail: shenye199594@stu.pku.edu.cn}~
\\
$^{1}$School of Physics, Peking University, No.5 Yiheyuan Rd, Beijing
		100871, P.R. China
}

\begin{document}

\maketitle

\begin{abstract}
    \vspace{5mm}
It is a hot topic nowadays that magnetic reconnection, as a physical process to release magnetic energy effectively, occurs in numerous complicated astrophysical systems. Since the magnetic reconnection is thought to occur frequently in the accretion flow around compact objects which induce strong gravitational field, it is now regarded to be a practical mechanism to extract energy from rotation black holes, which motivates people to consider how to describe the process of magnetic reconnection in a generally relativistic way. In this work, I try to explore the description of Sweet-Parker model, one of the most famous theoretical models of magnetic reconnection, in general relativity. I begin with revisiting the Sweet-Parker model in special relativity and reorganize the calculations in seven steps, whose generally relativistic forms are discussed. I propose in this work, from the general discussions and consequences of specific examples, that no property in Sweet-Parker model would be modified by spacetime curvature, which is opposite to the conclusions in previous work. However, on the contrary, observation in different rest frames may bring modifications. If the magnetic reconnection occurs not in the rest frame of observer, the observer would find out that the detected relation between the reconnection rate and Lundquist number or that between outflow speed and Alfv{\'e}n velocity are not the same as the detected relations if the magnetic reconnection occurs just in the rest frame of observer.

\end{abstract}


\newpage
\baselineskip 18pt

\tableofcontents  

\section{Introduction}
\label{sec:intro}

Magnetic reconnection, a physical process releasing magnetic energy effectively, now becomes one of the hottest topics in astrophysics. It generates the stellar flares \cite{substorm,2021SSRv..217...66Z} and drives the mass ejection in the solar corona \cite{Lin-Forbes-2000} or the coronae above the accretion disks of compact objects \cite{Yuan2009}. It also provides a possible explanations about the episodic jet flow \cite{Aimar2023,Ripperda2020}. Theoretical models were boosted to described the magnetic reconnection in slow \cite{SP1,SP2} and fast \cite{Petschek,Liu2017} reconnection rate under the scheme of magetohydrodynamics (MHD). To date, with the develoment of generally relativistic particle-in-cell \cite{Sironi:2014jfa,Comisso:2023ygd,Vay:2016hug} and magnetohydrodynamic \cite{Porth:2016rfi,White:2015omx} simulations, it is currently more convenient for people to numerically predict the occurrences of magnetic reconnection and their signatures from observation in complex astrophysical systems \cite{Jia:2023iup,Davelaar:2023dhl} in recent works.

It is widely believed that the magnetic reconnection occurs frequently in accretion disks, especially on the equatorial plane and near the central compact object where the magnetic field lines are highly curved and twisted \cite{Yuan2024-1,Yuan2024-2}. In this sense, extracting energy via magnetic reconnection, as a much more practical ignition of Penrose process than particle fission, from a rotating black hole (BH) now obtains great attentions. The feasibility of energy extraction from a Kerr BH were checked in different scenarios \cite{KA2008,CA2021,Work0}, followed by the analyses on different types of spacetime \cite{Carleo:2022qlv,Wei:2022jbi,Liu:2022qnr,Wang:2022qmg}. Most of the works about energy extraction adopted the Sweet-Parker model \cite{SP1,SP2}, the first analytical model to describe the magnetic reconnection. Though the extremely slow reconnection rate concluded from this model could not explain various astrophysical phenomena \cite{yamada2009}, it is simple and analyzable when applied in many complicated astrophysical scenarios as pioneering studies. Moreover, extracting energy from a rotating BH requires the occurence of magnetic reconnection within the ergo sphere where the gravitational effect is significant, which motivates people to analyze the Sweet-Parker model in a generally relativitic (GR) way.

In Ref.~\cite{Lyubarsky2006}, Sweet-Parker model, so well as the Petchek model \cite{Petschek}, was promoted to a specially relativistic (SR) model which can describe the magnetic reconnection in the limit of high magnetization. A GR description of Sweet-Parker model was proposed by Asenjo and Comisso in Ref.~\cite{CA2017}, just after their work about magnetic connection in GR \cite{Asenjo:2017rhf}. However, as an exploratory study, some problems in this work are pending. For example, equations, cast into a general form of $\nabla_{\nu}S^{\mu\nu}=0$, are all projected onto the frame of zero-angular-momentum observers (ZAMOs). This projection is actually cumbersome and it is hard to extend the description to the rest frames other than ZAMOs. What's more, to describe a physical process in general relativity, it is of vital importance to determine the laboratory, in the rest frame of which the process happens, and the observer, by which the process is detected. It seems that in Ref.~\cite{CA2017} ZAMOs was chosen to be the laboratory. However, the quasi-stationary condition they chose ($\partial_t\approx 0$) was in conflict with their choice of laboratory. What is worse, Ref.~\cite{CA2017} never considered the case that the magnetic reconnection process is detected not in the rest frame of laboratory. Additionally, in Sweet-Parker model, derivatives would be substituted by finite differences over the current sheet. However, this substitutions seemed not to be chosen in correct forms in Ref.~\cite{CA2017}.

In this work, I aim to explore how to describe the Sweet-Parker model in general relativity. I will start from revisiting the Sweet-Parker model, including its basic equations and approximations. The calculations in Sweet-Parker model will be reorganized in seven steps. Then I will try to discuss the Sweet-Parker model reconstructed on the basis of general relativity and give the GR forms of seven-steps calculations. Vitally, general forms of quasi-stationary condition and the substitutions in the Sweet-Parker model will be suggested. I will state my opinion, from general discussions and calculations in specific cases, that no property of magnetic reconnection would be modified significantly by spacetime curvature, which is opposite to the conclusions in Ref.~\cite{CA2017}. But the detected properties in the view of observer, on the contrary, would probably be changed. Furthermore, I will calculate the modifications of properties in Sweet-Parker model in the case that the observer is not at rest with respect to the laboratory.

The remaining parts of the paper are organized as follows. A concise recap of Sweet-Parker model will be shown in Sect.~\ref{sec:recap}. A general discussions about describing the Sweet-Parker model in GR will be introduced in Sect.~\ref{sec:GR}, including basic setups in Sect.~\ref{sec:setup}, calculations in GR forms in Sect.~\ref{sec:calculation} and the effect of observation in Sect.~\ref{sec:obs}. Specific examples will be demostrated in Sect.~\ref{sec:app}, including the dicussions about modifications of properties made by graviational effect and those caused by observation. I will summerize this work briefly in Sect.~\ref{sec:sum}. Units such that $G=M=c=1$ are adopted throughout the paper, without loss of generality.

\section{Recap of Sweet-Parker model}
\label{sec:recap}

\begin{figure}
    \centering
    \includegraphics[width=\textwidth]{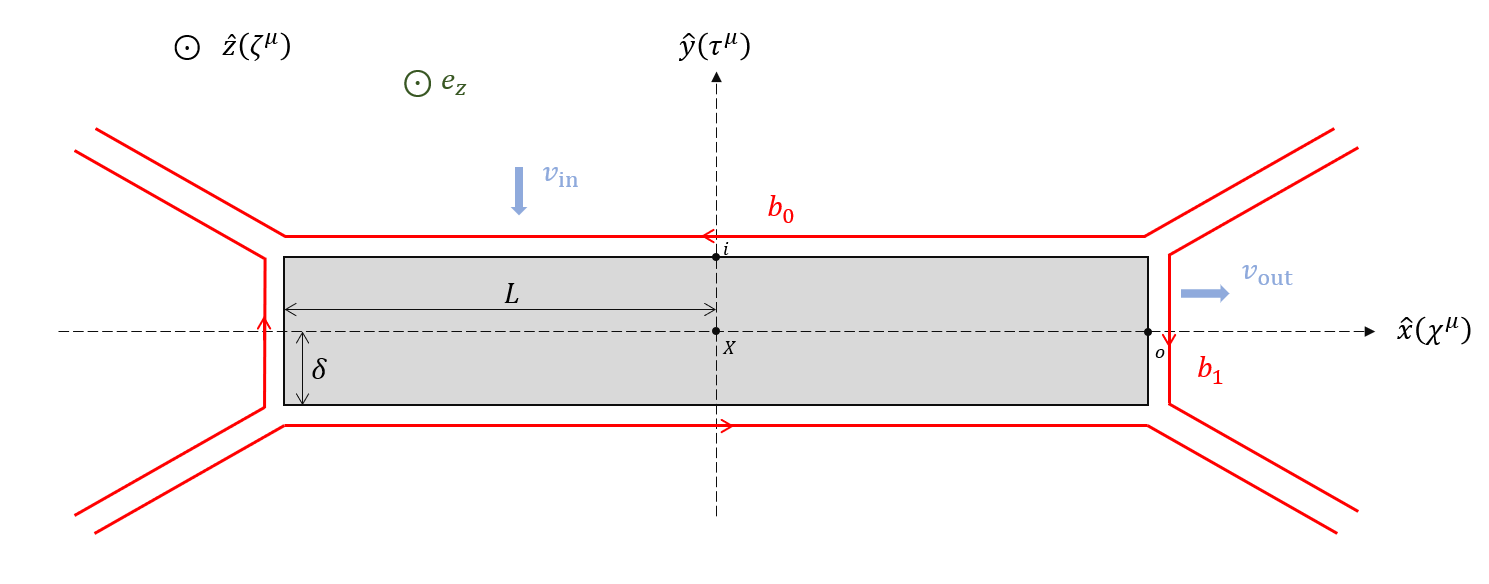}
    \caption{Schematic diagram of Sweet-Paker model, with red lines representing the magnetic field and the gray square representing the current sheet. Here $X$ denotes the center of current sheet while $i$ and $o$ denotes the surfaces of current sheet upstream and downstream.}
    \label{fig:pic}
\end{figure}

Calculations of Sweet-Parker model in SR are revisited in this section. The schematic diagram of the Sweet-Parker model is depicted on Fig.~\ref{fig:pic}, where the current sheet is posited on $\hat{x}-\hat{y}$ plane. The reconnection point, surfaces of current sheet upstream and downstream are denoted by $X$, $i$ and $o$ respectively. Magnetic strengths upstream and downstream are respectively set to be $\Vec{B}_i=b_0\hat{x}$ and $\Vec{B}_o=b_1\hat{y}$. The current sheet is assumed to be narrow ($\delta\ll L$) and quasi-two-dimensional which restricts the magnetic field to locate on the $\hat{x}-\hat{y}$ plane only while the electric field to direct along $\hat{z}$. Plasma flows into the current sheet along $\hat{y}$ with speed $v_{\rm in}$ and flows out along $\hat{x}$ with speed $v_{\rm out}$. Current exists within the current sheet only. 

Sweet-Parker model described the magnetic reconnection under the scheme of MHD. Four groups of equations should be taken into account:
\begin{equation}
    \begin{split}
        \text{\bf{ mass conservation:}}&~~\nabla\cdot\left(\gamma\Vec{v}\right)=0 \\
        \text{\bf{ momentum~conservation:}}&~~\nabla\cdot\left(\omega\gamma^2\Vec{v}\Vec{v}\right)+\nabla p=\Vec{J}\times\Vec{B} \\
        \text{\bf{ Ohm's~law:}}&~~\Vec{E}+\Vec{v}\times\Vec{B}=\eta\Vec{J} \\
        \text{\bf{ Maxwell's~equations:}}&~~\nabla\cdot\Vec{E}=0~,~\nabla\times\Vec{E}=0~,~\nabla\cdot\Vec{B}=0~,~\nabla\times\Vec{B}=\Vec{J} 
    \end{split}
    \label{eq:MHD}
\end{equation}
where the quasi-stationary condition \cite{CA2017}
\begin{equation}
    \partial_tq\sim 0
    \label{eq:steady}
\end{equation}
with $q$ an arbitrary physical quantity, is already considered. Here $\eta$ is the electric resistivity and $\omega$ is the enthalpy of plasma. Generally regarded, reconnection happens in a local scale (respect to the background), within which both of $\eta$ and $\omega$ are treated to reach uniformity \cite{Liu2017}. For the magnetic reconnection occurs in solar system, the local scale just means $\delta\ll L \ll r_{\odot}$, where $r_{\odot}$ denotes the radius of sun. Plasma in this local scale is neutral. There are totally 15 equations in Eq.~\eqref{eq:MHD}. However, many equations are trivial (for they would conclude $0=0$) or dispensable (for they consider the variations along $\hat{z}$) because it is a quasi-two-dimensional model. Derivatives in Eq.~\eqref{eq:MHD} can be substituted by finite differences in the form of \cite{Lyubarsky2006}:
\begin{equation}
    \partial_xq\sim\frac{q\big|_o-q\big|_X}{L}~~~,~~~\partial_yq\sim\frac{q\big|_i-q\big|_X}{\delta}
    \label{eq:L_delta}
\end{equation}
as a general approximation in order to handle the equations within the local scale for simplicity.

It is necessary to declare here that, in the following statement, the capital letters denote the components of vectors while the lowercase letters denote their values. For example, $B_x$ denotes the $\hat{x}$ component of magnetic field while $b_0$ denotes its strength upstream. Formally, we have, for example, $B_x\big|_i=b_0$ and $B_x\big|_X=B_x\big|_o=0$.

The main purposes of Sweet-Parker model are to find out the properties of outflow speed $v_{\rm out}$ and local reconnection rate $R$ which is defined to be the Mach number of plasma upstream. I reorganize the calculations into the following seven steps:
\begin{itemize}
\item First, as electric field directs $\hat{z}$ axis only, the null curl of electric field becomes:
\begin{equation}
    \partial_x E_z = \partial_y E_z =0~~\rightarrow~~ E_z=const\equiv e_z
    \label{eq:s1}
\end{equation}
These two equations give a constant electric field within this local scale (inside or outside the current sheet).

\item Second, the null divergence of the electric field gives:
\begin{equation}
    \partial_x B_x+\partial_y B_y=0~~\rightarrow~~ \frac{b_0}{L}\simeq\frac{b_1}{\delta}
    \label{eq:s2}
\end{equation}
from which one can find out $b_0\gg b_1$.

\item Third, Ohm's law upstream and inside the current sheet gives:
\begin{equation}
    \begin{split}
        {\rm upstream:}&~~e_z=v_{\rm in}b_0 \\
        {\rm current~sheet:}&~~J_z=\frac{e_z}{\eta}\equiv j_z
    \end{split}
    \label{eq:s3}
\end{equation}
Both of the equations above are the $\hat{z}$-component of Ohm's law whose other two components are trivial.

\item Fourth, the Amp{\`e}re's law surrounding the current sheet, namely the $\hat{z}$-component of $\nabla\times\Vec{B}=\Vec{J}$ over the current sheet, gives:
\begin{equation}
    \partial_x B_y - \partial_y B_x = J_z~~\rightarrow~~ \frac{j_z}{b_0} \simeq \frac{1}{\delta}
    \label{eq:s4}
\end{equation}
The derived equation on the right side applies the approximations $b_0\gg b_1$ and $L\gg\delta$. Combining the derived equations above, one gets:
\begin{equation}
    v_{\rm in}\simeq \frac{\eta}{\delta}
    \label{eq:vin}
\end{equation}

\item Fifth, mass conservation gives:
\begin{equation}
    \partial_x \gamma V_x = \partial_y \gamma V_y~~\rightarrow~~ \frac{\gamma_{\rm in}v_{\rm in}}{\gamma_{\rm out}v_{\rm out}} \simeq \frac{\delta}{L}
    \label{eq:s5}
\end{equation}
where $\gamma_{\rm in}$ and $\gamma_{\rm out}$ are Lorentz factors of inflow and outflow plasma respectively. The equation on the right is got by considering the conservation between $i$ and $o$. From Eq.~\eqref{eq:s5}, one concludes $v_{\rm in}\ll v_{\rm out}$ and subsequently $v_{\rm in}\ll 1$. It is required in Sweet-Parker model that the inflow plasma threading the surfaces of current sheet upstream can only be ejected from the surfaces of current sheet downstream, which consequently restrict the local reconnection rate \cite{Petschek,yamada2009}. 

\item Sixth, incompressibility of current sheet, coming from the momentum conservation along $\hat{y}$ axis, gives:
\begin{equation}
    \partial_y\left(\omega\gamma^2V_y^2\right)+\partial_y p=J_zB_x~~\rightarrow~~p_X\simeq \frac{b_0^2}{2}
    \label{eq:s6}
\end{equation}
The approximation $v_{\rm in}\ll 1$ concluded from Eq.~\eqref{eq:s5} and the relation we got in Eq.~\eqref{eq:s4} are applied. Here $p_X$ denotes the pressure of fluid inside the current sheet.

\item Seventh, the conservation of momentum along $\hat{x}$ axis shows:
\begin{equation}
    \partial_x\left(\omega\gamma^2V_x^2\right)+\partial_x p=-J_zB_y~~\rightarrow~~ \frac{1}{L}\left(\omega\gamma_{\rm out}^2v_{\rm out}^2-p_X\right)\simeq \frac{b_0^2}{2}
    \label{eq:s7}
\end{equation}
from which one concludes that:
\begin{equation}
    v_{\rm out}\simeq v_{\rm A}=\sqrt{\frac{\sigma_0}{1+\sigma_0}}
    \label{eq:vout}
\end{equation}
Here $\sigma_0=b_0^2/\omega$ is the local magnetization of plasma while $v_{\rm A}$ is the local Alfv{\'e}n velocity. The relations we got from Eq.~\eqref{eq:s2}, \eqref{eq:s4} and \eqref{eq:s6} are used to get Eq.~\eqref{eq:vout}. 

Local reconnection rate, in mildly relativistic case ($\gamma_{\rm in}\simeq 1$, $\gamma_{\rm out}\simeq 1$), satisfies
\begin{equation}
    R\equiv \frac{v_{\rm in}}{v_{\rm A}}\simeq \frac{\delta}{L} \simeq \frac{\eta}{\delta v_{\rm A}} \simeq S^{-1/2}
    \label{eq:R}
\end{equation}
based on the relations concluded from Eq.~\eqref{eq:s1}--\eqref{eq:s7}. Here $S\equiv Lv_{\rm A}/\eta$ is the Lundquist number of plasma which is large in most astrophysical system \cite{yamada2009}, resulting in the slow reconnection rate in Sweet-Parker model.
\end{itemize}

In the following sections, I will try to explore how to concisely express equations in the seven steps above in GR and how the equations in GR maps to their SR versions.

\section{Sweet-Parker model in GR: general discussions} 
\label{sec:GR}

\subsection{Basic setups} 
\label{sec:setup}

Basic equations of MHD in GR are:
\begin{equation}
    \begin{split}
        \text{\bf{ mass conservation:}}&~~\nabla_{\mu}(\rho u^{\mu})=0 \\
        \text{\bf{ energy-momentum~conservation:}}&~~\nabla_{\mu}T^{\mu\nu}=F^{\nu\mu}J_{\mu} \\
        \text{\bf{ Ohm's~law:}}&~~F^{\mu\nu}u_{\nu}=\eta\left(J^{\mu}+J_{\nu}u^{\nu}u^{\mu}\right) \\
        \text{\bf{ Maxwell's~equations:}}&~~\nabla_{\nu}F^{\mu\nu}=J^{\mu}~~,~~\nabla_{\nu}F^{\ast\mu\nu}=0
    \end{split}
    \label{eq:GR-MHD1}
\end{equation}
with $T^{\mu\nu}=w u^{\mu}u^{\nu}+pg^{\mu\nu}$ the stress-energy tensor. The plasma is considered to be neutral overall so that the second term in the right-hand side of Ohm's law reduces to zero. In local scale, $\rho$, $\omega$ and $\eta$ reaches uniformity \cite{Liu2017}. For the magnetic reconnection occurs near a BH, the local scale just means: $\delta\ll L\ll r_g$, where $r_g\equiv 1$ is the gravitational radius of the BH. 

As generally known, the covariant divergence for an arbitrary vector $A^{\mu}$ satisfies
\begin{equation}
    \nabla_{\mu}A^{\mu}=\frac{1}{\sqrt{-g}}\partial_{\mu}\left(\sqrt{-g}A^{\mu}\right)
\end{equation}
where $g$ is the determinant of metric tensor. While the covariant divergence for an arbitrary two dimentional tensor $G^{\mu\nu}$ satisfies 
\begin{equation}
    \nabla_{\mu}G^{\mu\nu}=\frac{1}{\sqrt{-g}}\partial_{\mu}\left(\sqrt{-g}G^{\mu\nu}\right)+\Gamma^{\mu}_{\kappa\lambda}G^{\lambda\kappa}
\end{equation}
where $\Gamma^{\mu}_{\kappa\lambda}$ are Christoffel symbols (affine connection). Then equations in Eq.~\eqref{eq:GR-MHD1} can be expressed as:
\begin{equation}
    \begin{split}
        \text{\bf{ mass conservation:}}&~~\partial_{\mu}(\sqrt{-g} u^{\mu})=0 \\
        \text{\bf{ energy-momentum~conservation:}}&~~\frac{1}{\sqrt{-g}}\partial_{\nu}\left(\sqrt{-g}\omega u^{\mu}u^{\nu}\right)+\partial^{\mu}p =
                                             F^{\mu\nu}J_{\nu}-\omega\Gamma^{\mu}_{\kappa\lambda}u^{\lambda}u^{\kappa} \\
        \text{\bf{ Ohm's~law:}}&~~F^{\mu\nu}u_{\nu}=\eta J^{\mu} \\
        \text{\bf{ Maxwell's~equations:}}&~~\partial_{\nu}\left(\sqrt{-g}F^{\mu\nu}\right)=\sqrt{-g}J^{\mu}~~
                                   ,~~\partial_{\nu}\left(\sqrt{-g}F^{\ast\mu\nu}\right)=0
    \end{split}
    \label{eq:GR-MHD2}
\end{equation}
The antisymmetries of $F^{\mu\nu}$ and $F^{\ast\mu\nu}$ were used here. One can check from Eq.~\eqref{eq:GR-MHD2} that if the first order derivatives of $g_{\mu\nu}$ are all neglected in local scale, the equations so well as the conclusions should have no difference to their SR versions introduced in Sect~\ref{sec:recap}.

For quantifying any physical process in GR, we have to specify the laboratory in the rest frame of which this process occurs and the observer who detects this process. For the magnetic reconnection as what Sweet-Parker model describes, current sheet should be posited on the spacelike hypersurface perpendicular to the 4-velocity of laboratory. In this sense, let us define:
\begin{equation}
    \begin{split}
        k^{\mu}&  \text{: 4-velocity of laboratory}\\
        n^{\mu}&  \text{: 4-velocity of observer}\\
        \chi^{\mu}&  \text{: spacelike unit vector along which the magnetic field is posited upstream}\\
        \tau^{\mu}&  \text{: spacelike unit vector along which the magnetic field is posited downstream}\\
        \zeta^{\mu}&  \text{: spacelike unit vector along which the electric field is posited}
    \end{split}
    \label{eq:tetrad}
\end{equation}
It is required that $k^{\mu}$ is perpendicular to $\chi^{\mu}$, $\tau^{\mu}$ and $\zeta^{\mu}$. Namely, for example, $\chi_{\mu}k^{\mu}=0$. Actually, the $\chi^{\mu}$, $\tau^{\mu}$ and $\zeta^{\mu}$ map to the $\hat{x}$, $\hat{y}$ and $\hat{z}$ axes in Fig.~\ref{fig:pic} and are perpendicular to each other. With the vectors defined in Eq.~\eqref{eq:tetrad}, the electric and magnetic field in the local scale observed in laboratory could be expressed as:
\begin{equation}
    E_{(k)}^{\mu}=E_z\zeta^{\mu}~~~~,~~~~B_{(k)}^{\mu}=B_x\chi^{\mu}+B_y\tau^{\mu}
    \label{eq:field}
\end{equation}
for: 
\begin{equation}
    \begin{split}
        \text{upstream:~}&B_x\big|_i=b_0,~~B_y\big|_i=0 \\
        \text{downstream:~}&B_x\big|_o=0,~~B_y\big|_o=b_1 \\
        \text{current sheet:~}&B_x\big|_X=B_y\big|_X=0
    \end{split}
    \label{eq:field2}
\end{equation} 
The Maxwell's tensors are then:
\begin{equation}
    \begin{split}
        F^{\mu\nu}&=E_{(k)}^{[\mu}k^{\nu]}+\epsilon^{\mu\nu\kappa\lambda}k_{\kappa}B^{(k)}_{\lambda} \\
        F^{\ast\mu\nu}&=B_{(k)}^{[\mu}k^{\nu]}+\epsilon^{\mu\nu\kappa\lambda}k_{\kappa}E^{(k)}_{\lambda}
    \end{split}
    \label{eq:Maxwell}
\end{equation}
Here $\epsilon^{\mu\nu\kappa\lambda}=-\left(-g\right)^{-1/2}[\mu\nu\kappa\lambda]$, with $[\mu\nu\kappa\lambda]$ the 4D Levi-Civita tensor. The 4-velocity of plasma should be:
\begin{equation}
    u^{\mu}=\gamma\left(k^{\mu}+V_x\chi^{\mu}+V_y\tau^{\mu}\right)
    \label{eq:velocity}
\end{equation}
for
\begin{equation}
    \begin{split}
        \text{upstream:~}&\gamma\big|_i\equiv\gamma_{\rm in}=\left(1-v_{\rm in}\right)^{-1/2},~~V_x\big|_i=0,~~V_y\big|_i=-v_{\rm in} \\
        \text{downstream:~}&\gamma\big|_o\equiv\gamma_{\rm in}=\left(1-v_{\rm out}\right)^{-1/2},~~V_x\big|_o=v_{\rm out},~~V_y\big|_o=0 \\
        \text{current sheet:~}&\gamma\big|_X=1,~~V_x\big|_X=V_y\big|_X=0 
    \end{split}
    \label{eq:velocity2}
\end{equation}
Here $v_{\rm in}$\footnote{The negative sign tells the flowing direction for we choose the top right quarter in Fig.~\ref{fig:pic} to describe.} and $v_{\rm out}$ are speeds of inflow and outflow plasma while $\gamma_{\rm in}$ and $\gamma_{\rm out}$ are their Lorentz factor respectively, observed in the rest frame of laboratory. 

After defining $k^{\mu}$, one can express the quasi-stationary condition in Eq.~\eqref{eq:steady} as 
\begin{equation}
    \partial_tq\sim 0~~~\xrightleftharpoons[\text{SR form}]{\text{GR form}}~~~\mathcal{L}_{\hat{k}}q\approx k^{\mu}\partial_{\mu}q\sim 0
    \label{eq:steady_GR}
\end{equation}
where $q$ is any physical quantity and $\mathcal{L}_{\hat{k}}$ is the Lee derivative along $k^{\mu}$. Recall that all the derivatives in Eq.~\eqref{eq:GR-MHD2} consider the variations in local scale, within which $\partial_{\nu}k^{\mu}\simeq 0$ is generally satisfied such that $\mathcal{L}_{\hat{k}}\approx k^{\mu}\partial_{\mu}$. Similarly, in GR, with the definition of $\chi^{\mu}$ and $\tau^{\mu}$, the substitutions in Eq.~\eqref{eq:L_delta} should be in the form of:
\begin{equation}
    \begin{split}
        \partial_xq\sim\frac{q\big|_o-q\big|_X}{L}~~~\xrightleftharpoons[\text{SR form}]{\text{GR form}}~~~\mathcal{L}_{\hat{\chi}}q\approx \chi^{\mu}\partial_{\mu}q\sim \frac{q\big|_o-q\big|_X}{L} \\
        ~~~~~~~~~~~~~~~~~~~~~~~~~~\\
        \partial_yq\sim\frac{q\big|_i-q\big|_X}{\delta}~~~\xrightleftharpoons[\text{SR form}]{\text{GR form}}~~~\mathcal{L}_{\hat{\tau}}q\approx \tau^{\mu}\partial_{\mu}q\sim \frac{q\big|_i-q\big|_X}{L}
    \end{split}
    \label{eq:L_delta_GR}
\end{equation}
Those are different from the approximations made in Ref.~\cite{CA2017} and we will see they resulting in different conclusions in Sect.~\ref{sec:zamo} in details.

\subsection{Calculations in Sweet-Parker model}
\label{sec:calculation}

Now one could list the seven steps (as what Sect.~\ref{sec:recap} described) of Sweet-Parker model in GR after all the preparations introduced in Sect.~\ref{sec:setup} were made:
\begin{itemize}
\item First, null curl of electric field, namely the spatial components of $\nabla_{\nu}F^{\ast\mu\nu}=0$:
\begin{equation}
    \partial_x E_z=\partial_y E_z=0~~~\xrightleftharpoons[\text{SR form}]{\text{GR form}}~~~
    \tau_{\mu}\partial_{\nu}\sqrt{-g}F^{\ast \mu\nu}=\chi_{\mu}\partial_{\nu}\sqrt{-g}F^{\ast \mu\nu}=0
    \label{eq:step1}
\end{equation}
The contraction with $\zeta^{\mu}$ gives a trivial equation.

\item Second, null divergence of magnetic field, namely the time component (along $k^{\mu}$ specifically) of $\nabla_{\nu}F^{\ast\mu\nu}=0$:
\begin{equation}
    \partial_xB_x+\partial_yB_y=0~~~\xrightleftharpoons[\text{SR form}]{\text{GR form}}~~~k_{\mu}\partial_{\nu}\sqrt{-g}F^{\ast \mu\nu}=0
    \label{eq:step2}
\end{equation}

\item Third, Ohm's law upstream and inside the current sheet:
\begin{equation}
    \begin{split}
        {\rm upstream:}&~E_z=-V_yB_x \\
        {\rm current~sheet:}&~E_z=\eta J_z
    \end{split}
    ~~~\xrightleftharpoons[\text{SR form}]{\text{GR form}}~~~
    \begin{split}
        &\zeta_{\mu}F^{\mu\nu}u_{\nu}=0 \\
        &\zeta_{\mu}F^{\mu\nu}u_{\nu}=\eta J^{\mu}\zeta_{\mu}
    \end{split}
    \label{eq:step3}
\end{equation}

\item Fourth, Amp{\`e}re's law surrounding the current sheet, namely the spatial components of $\nabla_{\nu}F^{\mu\nu}=J^{\mu}$:
\begin{equation}
    \partial_xB_y-\partial_yB_x=J_z~~~\xrightleftharpoons[\text{SR form}]{\text{GR form}}~~~
    \zeta_{\mu}\partial_{\nu}\sqrt{-g}F^{\mu\nu}=\sqrt{-g}J^{\mu}\zeta_{\mu}
    \label{eq:step4}
\end{equation}
The other two spatial components provide trivial equations.

\item Fifth, mass conservation:
\begin{equation}
    \partial_x\gamma V_x+\partial_y\gamma V_y=0~~~\xrightleftharpoons[\text{SR form}]{\text{GR form}}~~~\partial_{\mu}\sqrt{-g}u^{\mu}=0
    \label{eq:step5}
\end{equation}

\item Sixth, incompressibility of current sheet, coming from the conservation of energy-momentum along the direction of $\hat{y}$ axis (along $\tau^{\mu}$ in GR):
\begin{equation}
    \partial_y\left(\omega\gamma^2V_y^2\right)+\partial_y p=J_zB_x~~~\xrightleftharpoons[\text{SR form}]{\text{GR form}}~~~
    \frac{1}{\sqrt{-g}}\tau_{\mu}\partial_{\nu}\left(\sqrt{-g}\omega u^{\mu}u^{\nu}\right)+
    \tau_{\mu}\partial^{\mu}p=\tau_{\mu}F^{\mu\nu}J_{\nu}-\omega\tau_{\mu}\Gamma^{\mu}_{\kappa\lambda}u^{\kappa}u^{\lambda}
    \label{eq:step6}
\end{equation}

\item Seventh, the conservation of energy-momentum along the direction of $\hat{x}$ axis (along $\chi^{\mu}$ in GR):
\begin{equation}
    \partial_x\left(\omega\gamma^2v_x^2\right)+\partial_x p=-J_zB_y~~~\xrightleftharpoons[\text{SR form}]{\text{GR form}}~~~
    \frac{1}{\sqrt{-g}}\chi_{\mu}\partial_{\nu}\left(\sqrt{-g}\omega u^{\mu}u^{\nu}\right)+\chi_{\mu}\partial^{\mu}p
    =\chi_{\mu}F^{\mu\nu}J_{\nu}-\omega\chi_{\mu}\Gamma^{\mu}_{\kappa\lambda}u^{\kappa}u^{\lambda}
    \label{eq:step7}
\end{equation}
\end{itemize}

For any specific laboratory with the current sheet posited at rest on any plane arbitrarily, one would only need to utilize the specific forms of $k^{\mu}$, $\chi^{\mu}$, $\tau^{\mu}$ and $\zeta^{\mu}$ and duplicate the calculations listed from Eq.~\eqref{eq:step1} to Eq.~\eqref{eq:step7} for getting the description of Sweet-Parker model in GR. We will see some examples exhibited in Sect.~\ref{sec:app}.

\subsection{Observations}
\label{sec:obs}

Consider that the reconnection occurs in some laboratory while the reconnection rate and outflow speed are detected to obey the relations in Eq.~\eqref{eq:vout} and \eqref{eq:R} respectively in this laboratory, neglecting the probable perturbations from gravity since they are actually infinitesimal (which will be seen in Sect.~\ref{sec:app} in specific examples). When the reconnection rate and outflow speed are detected by an observer, the relations the observer concludes would become different. Here, the resistivity $\eta$ and enthalpy $\omega$ are four dimensional scalars such that they are invariant. While other quantities, the observed inflow and outflow speed $v_{\rm in/out,obs}$, the observed length of current sheet $L_{\rm obs}$ and the observed local Alfv{\'e}n velocity $v_{\rm A,obs}$, are all different from the values detected in the laboratory. 
\begin{itemize}
\item {\bf Observed inflow and outflow speed} 

For a general test particle with 4-velocity $m^{\mu}$, its 3-velocity measured by $n^{\mu}$ should be:
\begin{equation}
    v_m^{\mu}=-\frac{m^{\mu}+n_{\nu}m^{\nu}n^{\mu}}{n_{\nu}m^{\nu}}
\end{equation}
Assume the case that the observer is capable of measuring the velocities of the laboratory, inflow and outflow plasma. In the view of $n^{\mu}$, the magnitude of inflow and outflow speeds, with respect to the laboratory, should be:
\begin{equation}
    v_{\rm in/out,obs}=\left|-\frac{u^{\mu}\big|_{i/o}+n_{\nu}u^{\nu}\big|_{i/o} n^{\mu}}{n_{\nu}u^{\nu}\big|_{i/o}}+\frac{k^{\mu}+n_{\nu}k^{\nu}n^{\mu}}{n_{\nu}k^{\nu}}\right|
    \label{eq:v_obs}
\end{equation}
Here $\left|A^{\mu}\right|$ means calculating the length of 4-vector $A^{\mu}$.

\item {\bf Observed length of current sheet} 

The unit spacelike vector $\chi^{\mu}$, which is perpendicular to $k^{\mu}$ and along which the current sheet is posited, can be expressed as a linear combination of $n^{\mu}$ and another unit spacelike vector $\chi'^{\mu}$ perpendicular to $n^{\mu}$:
\begin{equation}
    \chi^{\mu}=Tn^{\mu}+F\chi'^{\mu}~~\rightarrow~~\chi'^{\mu}=\frac{1}{F}\chi^{\mu}-\frac{T}{F}n^{\mu}
\end{equation}
It is obvious that $F\neq 0$. The length of current sheet in the view of $n^{\mu}$ should be:
\begin{equation}
    L_{\rm obs}=\frac{L}{F}=\frac{L}{\left|\chi^{\mu}+n_{\nu}\chi^{\nu}n^{\mu}\right|}
    \label{eq:L_obs}
\end{equation}
where $L$ is just the length of current sheet viewed in the rest frame of laboratory.

\item {\bf Observed Alfv{\'e}n velocity}

The magnetic field in the view of $n^{\mu}$ should be:
\[
    B_{(n)}^{j}=-n_{\nu}F^{\ast j\nu}
\]
The observed magnetization and observed Alfv{\'e}n velovity upstream should be:
\begin{equation}
    \sigma_{\rm 0,obs}=\frac{\left(-n_{\nu}F^{\ast j\nu}\big|_i\right)^2}{\omega}~~~,~~~v_{\rm A,obs}=\sqrt{\frac{\sigma_{\rm 0,obs}}{1+\sigma_{\rm 0,obs}}}
    \label{eq:vA_obs}
\end{equation}
The proper enthalpy of plasma $\omega$ is a four dimensional scalar.

\item {\bf Observed relations}

With $v_{\rm in/out}$, $L_{\rm obs}$ and $v_{\rm A,obs}$, one can find out the relations:
\begin{equation}
    \frac{R_{\rm obs}}{S_{\rm obs}^{-1/2}}=\frac{v_{\rm in,obs}/v_{\rm A,obs}}{\left(L_{\rm obs}v_{\rm A,obs}/\eta\right)^{-1/2}}~~~,~~~
    \frac{v_{\rm out,obs}}{v_{\rm A,obs}}\simeq
    \begin{cases}
        &v_{\rm out,obs}/\sigma_{\rm 0,obs}^{1/2}~~~~\text{LM} \\
        &v_{\rm out,obs}/v_{\rm A}~~~~~~~~\text{HM}
    \end{cases}
    \label{eq:rela_obs}
\end{equation}
which do not equal one generally. Here $R_{\rm obs}$ and $S_{\rm obs}$ are the reconnection rate and Lundquist number in the view of $n^{\mu}$. The low magnetization limit ($\sigma_0\ll 1$, $\sigma_{\rm 0,obs}\ll 1$) is denoted to be "LM" while high magnetization limit ($\sigma_0\gg 1$, $\sigma_{\rm 0,obs}\gg 1$) is denoted to be "HM". The outflow speed, viewed in the laboratory, satisfies $v_{\rm out}\ll 1$ in the low magnetization limit while $v_{\rm out}\simeq 1$ in the high magnetization limit.
\end{itemize}

To consider the modifications of relations induced by observations, one could just utilize the equations from Eq.~\eqref{eq:v_obs} to Eq.~\eqref{eq:rela_obs} after determining laboratory, observer and how the current sheet is posited.

\section{Sweet-Parker model in GR: examples}
\label{sec:app}

\subsection{Metric and rest frames}
\label{sec:frames}

Let us focus on the stationary, axisymmetric spacetime, whose line elements could be written in 3+1 formalism as \cite{MacDonald:1982zz}:
\begin{equation}
    ds^2=-\alpha^2dt^2+\sum_{i=1}^{3}h_{i}^2\left(dx^i-\omega^{i}dt\right)^2
    \label{eq:element}
\end{equation}
with $\alpha$ the lapse function, $h_i$ the scale factors of the coordinates $x^i$ and $\omega^i$ the velocity corresponding to a frame dragging. Specifically, for the Kerr metric in Boyer-Lindquist (BL) coordinates $(t,r,\theta,\phi)$, we have \cite{CA2017}:
\begin{equation}
    \alpha=\sqrt{\frac{\Delta\Sigma}{A}},~~h_r=\sqrt{\frac{\Sigma}{\Delta}},~~h_{\theta}=\sqrt{\Sigma},~~h_{\phi}=\sqrt{\frac{A}{\Sigma}}\sin\theta,~~
    \omega^r=\omega^{\theta}=0,~~\omega^{\phi}=\frac{2ar}{A},
\end{equation} 
where $\Delta=r^2-2r+a^2$, $\Sigma=r^2+a^2\cos^2\theta$ and $A=\left(r^2+a^2\right)^2-\Delta a^2\sin^2\theta$. The determinant of metric tensor satisfies: $\sqrt{-g}=\alpha h_1 h_2 h_3$. 

Two kinds of important rest frames would be taken into concerns in the following to be the rest frame of laboratory. The first one is the rest frame of ZAMOs, which can be defined via the normal tetrad:
\begin{equation}
    \hat{e}_{(t)}^{\mu}=\frac{1}{\alpha}\left(\partial_t^{\mu}+\omega^{\phi}\partial_{\phi}^{\mu}\right);~~
    \hat{e}_{(r)}^{\mu}=\frac{1}{h_r}\partial_{r}^{\mu},~~
    \hat{e}_{(\theta)}^{\mu}=\frac{1}{h_{\theta}}\partial_{\theta}^{\mu},~~
    \hat{e}_{(\phi)}^{\mu}=\frac{1}{h_{\phi}}\partial_{\phi}^{\mu}
    \label{eq:ZAMO}
\end{equation}
Upon the normal tetrad of ZAMOs, one can define the rest frame of fluid moving on the equatorial plane via the normal tetrad in the form of \cite{Work0}:
\begin{equation}
    \begin{split}
        e_{[0]}^{\mu}&=\hat{\gamma}_s \left[\hat{e}_{(t)}^{\mu}+\hat{v}_s^{(r)}\hat{e}_{(r)}^{\mu}+\hat{v}_s^{(\phi)}\hat{e}_{(\phi)}^{\mu}\right]; \\
        e_{[1]}^{\mu}&=\frac{1}{\hat{v}_s}\left[\hat{v}_s^{(\phi)}\hat{e}_{(r)}^{\mu}-\hat{v}_s^{(r)}\hat{e}_{(\phi)}^{\mu}\right],~~
        e_{[2]}^{\mu}=\hat{e}_{(\theta)}^{\mu}, \\
        e_{[3]}^{\mu}&=\hat{\gamma}_s\left[\hat{v}_s\hat{e}_{(t)}^{\mu}+\frac{\hat{v}_s^{(r)}}{\hat{v}_s}\hat{e}_{(r)}^{\mu}+\frac{\hat{v}_s^{(\phi)}}{\hat{v}_s}\hat{e}_{(\phi)}^{\mu}\right]
    \end{split}
    \label{eq:plasma}
\end{equation}
where $\hat{v}_s=\sqrt{\left(\hat{v}_s^{(r)}\right)^2+\left(\hat{v}_s^{(\phi)}\right)^2}$ is the speed of fluid with $\hat{v}_s^{(r)}$ and $\hat{v}_s^{(r)}$ the two components of 3-velocity measured by ZAMOs and $\hat{\gamma}_s$ is the Lorentz factor. One can see that, viewed by ZAMOs, $e_{[3]}^{\mu}$ and $e_{[1]}^{\mu}$ are respectively parallel and perpendicular to the moving direction of fluid.

Before showing the calculations for specific laboratories and observers, it is inevitable to declare some points and approximations which might be useful in the following discussions:
\begin{itemize}
\item First, derivatives in the equations above relate to the variations in local scale. So either 4-velocity of laboratory or that of observer are independent of the derivatives. For example, the operator $k^{\mu}\partial_{\nu}$ satisfies:
\begin{equation}
    k^{\mu}\partial_{\nu}=\partial_{\nu}k^{\mu}
\end{equation}
when acting on any tensor. Cases are the same for $\chi^{\mu}$, $\tau^{\mu}$ and $\zeta^{\mu}$. 

\item Second, since the local scale is tiny, if we choose, for example, $\chi^{\mu}=\hat{e}_{(\phi)}^{\mu}$, the operator $\chi^{\mu}\partial_{\mu}\sqrt{-g}$ could be approximated as:
\begin{equation}
    \chi^{\mu}\partial_{\mu}\sqrt{-g}=\partial_{\phi}\frac{1}{h_{\phi}}\bigg|_X \alpha h_rh_{\theta}h_{\phi} \simeq \partial_{\phi}\alpha h_rh_{\theta}
\end{equation}
when acting on any tensor. This approximation will help to simplify the resultant expressions in Sect.~\ref{sec:calculation} 

\item Third, as a temporary approximation, we make, for example,
\begin{equation}
    \partial_{r}\alpha h_{\theta}h_{\phi}\simeq 0~~\xrightarrow{\div \alpha h_rh_{\theta}h_{\phi}}~~\frac{1}{h_r}\partial_{r} \simeq 0
    \label{eq:temp}
\end{equation}
which is just a neglect of first order derivatives of $g_{\mu\nu}$. It is actually illegal since $\alpha$ and $h_i$ are dependent on $r$. However, whether or not the GR forms of equations in Eq.~\eqref{eq:step1}--\eqref{eq:step7} could map to their SR forms in Eq.~\eqref{eq:s1}--\eqref{eq:s7} would be conveniently seen under this temporary approximation. It is worth noticing that the derivation
\begin{equation}
    \partial_{\phi}\alpha h_{\theta}h_{\phi}\simeq 0~~\xrightarrow~~\frac{1}{h_r}\partial_{\phi} \simeq 0
\end{equation}
is always legal because $\alpha$ and $h_i$ are independent of $\phi$.

\item Fourth, when substituting the derivatives by finite differences based on Eq.~\eqref{eq:L_delta_GR}, to manage the terms with factors outside the derivatives, formally the terms like $f_1\tau^{\mu}\partial_{\mu}f_2$, one could make, for example between $i$ and $X$:
\begin{equation}
    f_1\tau^{\mu}\partial_{\mu}f_2 \simeq f_1\big|_{\varepsilon}\frac{f_2\big|_i-f_2\big|_X}{\delta}
    \label{eq:mean}
\end{equation}
with $\varepsilon$ some point between $i$ and $X$, on which the mean value theorem for finite integral is satisfied.
\end{itemize}

\subsection{ZAMOs laboratory with azimuthal current sheet}
\label{sec:zamo}

In this section, let us consider that the magnetic reconnection occurs in the rest frame of ZAMOs (called ZAMOs laboratory henceforth). The current sheet is posited azimuthally on equatorial plane while the $\hat{y}$ axis is posited radially. That is to say:
\begin{equation}
    k^{\mu}=\hat{e}_{(t)}^{\mu};~~\chi^{\mu}=\hat{e}_{(\phi)}^{\mu},~~\tau^{\mu}=\hat{e}_{(r)}^{\mu},~~\zeta^{\mu}=\hat{e}_{(\theta)}^{\mu}
    \label{eq:tetrad_z}
\end{equation} 
Lee derivatives along $k^{\mu}$, $\chi^{\mu}$ and $\tau^{\mu}$ should be:
\begin{equation}
    \mathcal{L}_{\hat{k}}\approx\frac{1}{\alpha}\partial_t+\frac{\omega^{\phi}}{\alpha}\partial_{\phi}
    \label{eq:steady_z}
\end{equation}
and
\begin{equation}
    \mathcal{L}_{\hat{\chi}}\approx\frac{1}{h_{\phi}}\partial_{\phi}~~,~~\mathcal{L}_{\hat{\tau}}\approx\frac{1}{h_r}\partial_{r}
    \label{eq:L_delta_z}
\end{equation}
It is the same as the first case described in Ref.~\cite{CA2017}. Now let us exhibit the specific expressions of Eq.~\eqref{eq:step1} - \eqref{eq:step7} by utilizing Eq.~\eqref{eq:tetrad_z}, \eqref{eq:steady_z} and \eqref{eq:L_delta_z}.
\begin{itemize}

\item Eq.~\eqref{eq:step1} gives:
\begin{equation}
    \begin{split}
        \partial_r\left(\alpha h_{\theta}h_{\phi}E_z\right)+\partial_t\left(h_rh_{\theta}h_{\phi}B_x\right)+\partial_{\phi}\left(h_rh_{\theta}h_{\phi}\omega^{\phi}B_x\right)=0~~&\xrightarrow{\div \alpha h_rh_{\theta}h_{\phi}}~~
        \frac{1}{h_r}\partial_{r}E_z+\left(\frac{1}{\alpha}\partial_t+\frac{\omega^{\phi}}{\alpha}\partial_{\phi}\right)B_x \simeq 0 \\
        \partial_{\phi}\left(\alpha h_rh_{\theta}E_z\right)+\partial_t\left(h_rh_{\theta}h_{\phi}B_y\right)+\partial_{\phi}\left(h_rh_{\theta}h_{\phi}\omega^{\phi}B_x\right)=0~~&\xrightarrow{\div \alpha h_rh_{\theta}h_{\phi}}~~
        \frac{1}{h_{\phi}}\partial_{\phi}E_z+\left(\frac{1}{\alpha}\partial_t+\frac{\omega^{\phi}}{\alpha}\partial_{\phi}\right)B_y \simeq 0
    \end{split}
    \label{eq:z_step1}
\end{equation}
Approximately, the derivatives of $B_x$ and $B_y$ in the right side equations could be removed by quasi-stationary condition in Eq.~\eqref{eq:steady_GR}. Equations on the right side conclude $E_z \simeq const \equiv e_z$ within the local scale, same as the conclusion of the SR version in this step.

\item Eq.~\eqref{eq:step2} gives:
\begin{equation}
    \partial_r \left(\alpha h_{\theta} h_{\phi} B_y\right) + \partial_{\phi}\left(\alpha h_r h_{\theta}B_x\right)=0
    ~~\xrightarrow{\div \alpha h_rh_{\theta}h_{\phi}}~~
    \frac{1}{h_r}\partial_{r}B_y+\frac{1}{h_{\phi}}\partial_{\phi}B_x \simeq 0
    \label{eq:z_step2}
\end{equation}
If one chooses the substitutions adopted in Ref.~\cite{CA2017}:
\begin{equation}
    L^{-1}\sim r^{-1}\partial_{\phi}~~,~~\delta^{-1}\sim \partial_{r}
    \label{eq:CA}
\end{equation} 
conclusion from the right side equation in Eq.~\eqref{eq:z_step2} would become
\begin{equation}
    \frac{b_1}{b_0} \simeq \frac{\delta}{L} \frac{r h_{r}}{h_{\phi}}
\end{equation}
which is identical to the Eq.~(7) in Ref.~\cite{CA2017}. However, utilizing the substitutions in Eq.~\eqref{eq:L_delta_GR} result in $b_1/b_0 \simeq \delta/L$, which is just the conclusion in SR. If one desires to consider the modification from gravitational effect, the relation between $b_0$ and $b_1$ should be concluded from the left side equation in Eq.~\eqref{eq:z_step2} as
\begin{equation}
    \frac{1}{h_r}\partial_r\left(\alpha h_{\theta} h_{\phi} B_y\right) + \frac{1}{h_{\phi}}\partial_{\phi}\left(\alpha h_{\theta} h_{\phi} B_x\right)=0
    ~~\rightarrow~~
    \frac{b_1}{b_0}\simeq \frac{\delta}{L} \frac{\left(\alpha h_{\theta}h_{\phi}\right)\big|_o}{\left(\alpha h_{\theta}h_{\phi}\right)\big|_i}\simeq \frac{\delta}{L}
\end{equation}
It is easily seen that the modification vanishes when $i$ and $o$ are extremely close.

\item Eq.~\eqref{eq:step3} gives:
\begin{equation}
    \begin{split}
        {\rm upstream:}&~\gamma_{\rm in}\left(e_z+v_{\rm in}b_0\right)=0 \\
        {\rm current~sheet:}&~J_z=\frac{E_z}{\eta}\simeq \frac{e_z}{\eta}\equiv j_z
    \end{split}
    \label{eq:z_step3}
\end{equation}

\item Eq.~\eqref{eq:step4} gives:
\begin{equation}
    \begin{split}
        &\left(\partial_t+\partial_{\phi}\omega^{\phi}\right)\left(h_rh_{\theta}h_{\phi}E_z\right)+\partial_{\phi}\left(\alpha h_rh_{\theta}B_y\right)-\partial_r\left(\alpha h_{\theta}h_{\phi}B_x\right)=\alpha h_rh_{\theta}h_{\phi} J_z \\
        \xrightarrow{\div \alpha h_rh_{\theta}h_{\phi}}~~& \left(\frac{1}{\alpha}\partial_t+\frac{\omega^{\phi}}{\alpha}\partial_{\phi}\right)E_z+\frac{1}{h_{\phi}}\partial_{\phi}B_y-\frac{1}{h_r}\partial_rB_x \simeq J_z
    \end{split}
    \label{eq:z_step4}
\end{equation}
The first term on the left side of the bottom equation in Eq.~\eqref{eq:z_step4} could be approximately removed by quasi-stationary condition in Eq.~\eqref{eq:steady_GR}. Neglecting the second term further on the left side (for $\delta\ll L$ and $b_1\ll b_0$) and making the substitution in Eq.~\eqref{eq:CA} would conclude:
\begin{equation}
    j_z \simeq -\frac{1}{h_r}\frac{b_0}{\delta}
\end{equation}
which is identical to the Eq.~(8) in Ref.~\cite{CA2017} (the negative sign tells the direction). While the substitutions in Eq.~\eqref{eq:L_delta_GR} provide:
\begin{equation}
    -\frac{1}{h_r}\partial_{r}\left(\alpha h_{\theta}h_{\phi}B_x\right)\simeq \alpha h_{\theta}h_{\phi}J_z~~\rightarrow~~
    \frac{j_z}{b_0}\simeq -\frac{1}{\delta}\frac{\left(\alpha h_{\theta}h_{\phi}\right)\big|_i}{\left(\alpha h_{\theta}h_{\phi}\right)\big|_{\varepsilon_1}}\simeq -\frac{1}{\delta}
    \label{eq:z_step4_con}
\end{equation}
Then Eq.~\eqref{eq:z_step3} and \eqref{eq:z_step4} conclude:
\begin{equation}
    v_{\rm in}\simeq \frac{\eta}{\delta}\frac{\left(\alpha h_{\theta}h_{\phi}\right)\big|_i}{\left(\alpha h_{\theta}h_{\phi}\right)\big|_{\varepsilon_1}}\simeq \frac{\eta}{\delta}
\end{equation}
where $\varepsilon_1$ is some point between $i$ and $X$ as argued in Eq.~\eqref{eq:mean}. This modification vanishes as well when $i$ and $X$ are close.

\item Eq.~\eqref{eq:step5} gives:
\begin{equation}
    \begin{split}
        & \partial_t\left(h_rh_{\theta}h_{\phi}\gamma\right)+\partial_r\left(\alpha h_{\theta}h_{\phi}\gamma V_y\right)+\partial_{\phi}\left[h_rh_{\theta}\gamma\left(h_{\phi}\omega^{\phi}+\alpha V_x\right)\right]=0  \\
        \xrightarrow{\div \alpha h_rh_{\theta}h_{\phi}}~~& \left(\frac{1}{\alpha}\partial_t+\frac{\omega^{\phi}}{\alpha}\partial_{\phi}\right)\gamma+\frac{1}{h_r}\partial_r\gamma V_y+\frac{1}{h_{\phi}}\partial_{\phi}\gamma V_x \simeq 0
    \end{split}
    \label{eq:z_step5}
\end{equation}
The 1st term on the left side of bottom equation in Eq.~\eqref{eq:z_step5} could be removed by quasi-stationary condition. Again making the substitution in Eq.~\eqref{eq:CA}, one gets:
\begin{equation}
    \frac{\gamma_{\rm in}v_{\rm in}}{\gamma_{\rm out}v_{\rm out}}\simeq \frac{\delta}{L}\frac{rh_r}{h_{\phi}}
\end{equation}
which is identical to the Eq.~(11) in Ref.~\cite{CA2017}. While applying substitutions in Eq.~\eqref{eq:L_delta_GR} concludes:
\begin{equation}
    \frac{1}{h_r}\partial_r\left(\alpha h_{\theta}h_{\phi}\gamma V_y\right)+\frac{1}{h_{\phi}}\partial_{\phi}\left(\alpha h_{\theta}h_{\phi}\gamma V_x\right) \simeq 0
    ~~\rightarrow~~\frac{\gamma_{\rm in}v_{\rm in}}{\gamma_{\rm out}v_{\rm out}}\simeq \frac{\delta}{L}\frac{\left(\alpha h_{\theta}h_{\phi}\right)\big|_o}{\left(\alpha h_{\theta}h_{\phi}\right)\big|_i}\simeq \frac{\delta}{L} 
\end{equation}
The modification vanishes as well when $i$ and $o$ are close.

\item Eq.~\eqref{eq:step6} gives:
\begin{equation}
    \begin{split}
        &\left[\partial_th_rh_{\theta}h_{\phi}+\partial_r\alpha h_{\theta}h_{\phi}V_y+
        \partial_{\phi}h_rh_{\theta}\left(h_{\phi}\omega^{\phi}+\alpha V_x\right)\right]\omega\gamma^2 V_y \\  
        &~~~~~~~~~~~~~~~~~~~~~~~~~~~~~~~~~~~~~~~~~~~+\alpha h_{\theta}h_{\phi}\partial_r p + \omega\Tilde{\mathcal{O}}_1\left(\Gamma\right)= 
        \alpha h_rh_{\theta}h_{\phi}B_x J_z \\
        ~~~~~~~~~~~~~~~~~~~~~~~ \\
        \xrightarrow{\div \alpha h_rh_{\theta}h_{\phi}}~~& \left[\left(\frac{1}{\alpha}\partial_t
        +\frac{\omega^{\phi}}{\alpha}\partial_{\phi}\right)+\frac{1}{h_r}\partial_r V_y+\frac{1}{h_{\phi}}\partial_{\phi} V_x\right]
        \omega\gamma^2 V_y +\frac{1}{h_r}\partial_r p + \omega\mathcal{O}\left(\Gamma\right) \simeq B_x J_z
    \end{split}
    \label{eq:z_step6}
\end{equation}
where:
\begin{equation}
    \mathcal{O}\left(\Gamma\right)\simeq h_r^{-3}\partial_r \ln\alpha
\end{equation}
under the approximation $V_y\ll 1$. Terms framed by brakets in Eq.~\eqref{eq:z_step6} should be regarded as operators acting on $\omega\gamma^2V_y$. On the left side of bottom equation in Eq.~\eqref{eq:z_step6}, the 1st term in the braket could be removed by quasi-stationary condition in Eq.~\eqref{eq:steady_GR}. Considering the variation between point $i$ and $X$ and utilizing the relation between $b_0$ and $j_z$ further, one gets:
\begin{equation}
    \frac{1}{h_r}\partial_rp + \frac{\omega\partial_r\ln\alpha}{h_r^3} \simeq -\frac{B_x}{\alpha h_rh_{\theta}h_{\phi}}\partial_r\left(\alpha h_{\theta}h_{\phi}B_x\right)~~\rightarrow~~
    \frac{p_X}{b_0^2/2}\simeq \frac{\left(\alpha^2h_{\theta}^2h_{\phi}^2\right)\bigg|_i}{\left(\alpha^2h_{\theta}^2h_{\phi}^2\right)\bigg|_{\varepsilon_2}}-
    \frac{\omega}{b_0^2/2}\frac{\ln\alpha\big|_i-\ln\alpha\big|_X}{h_r^2\big|_{\varepsilon_3}}\simeq 1
\end{equation}
where $\varepsilon_2$ and $\varepsilon_3$, which are generally different, are some points between $i$ and $X$. The the first term of the top equation in Eq.~\eqref{eq:z_step6} is removed for $V_y\ll 1$.

\item Eq.~\eqref{eq:step7} gives:
\begin{equation}
    \begin{split}
        & \left[\partial_th_rh_{\theta}h_{\phi}+\partial_r\alpha h_{\theta}h_{\phi}V_y+\partial_{\phi}h_rh_{\theta}\left(h_{\phi}\omega^{\phi}+\alpha V_x\right)\right]\omega\gamma^2 V_x +\alpha h_rh_{\theta}\partial_{\phi}p
        =-\alpha h_rh_{\theta}h_{\phi}B_yJ_z \\
        \xrightarrow{\div \alpha h_rh_{\theta}h_{\phi}}~~& \left[\left(\frac{1}{\alpha}\partial_t+\frac{\omega^{\phi}}{\alpha}\partial_{\phi}\right)
        +\frac{1}{h_r}\partial_rV_y+\frac{1}{h_{\phi}}\partial_{\phi}V_x\right]\omega\gamma^2V_x+\frac{1}{h_{\phi}}\partial_{\phi}p
        \simeq -B_y J_z 
    \end{split}
    \label{eq:z_step7}
\end{equation}
Term $\chi_{\mu}\Gamma^{\mu}_{\kappa\lambda}u^{\kappa}u^{\lambda}$ equals zero here. Terms framed by brakets should be regarded as operators acting on $\omega\gamma^2V_x$. On the left side of bottom equation, the first term in braket could be removed by quasi-stationary condition in Eq.~\eqref{eq:steady_GR}. Considering the variation between $X$ and $o$ and applying the relations among $b_0$, $b_1$ and $j_z$, one gets:
\begin{equation}
    v_{\rm out}\simeq \sqrt{\frac{\mathscr{D}\sigma_0}{1+\mathscr{D}\sigma_0}}
\end{equation}
where $\mathscr{D}$ is the factor of modification:
\begin{equation}
    \mathscr{D}\equiv \frac{1}{2}\left[\frac{\left(\alpha^2h_{\theta}^2h_{\phi}^2\right)\bigg|_i}{\left(\alpha^2h_{\theta}^2h_{\phi}^2\right)\bigg|_{\varepsilon_2}}-
    \frac{\omega}{b_0^2/2}\frac{\ln\alpha\big|_i-\ln\alpha\big|_X}{h_r^2\big|_{\varepsilon_3}} + 
    \frac{\left(\alpha h_{\theta}h_{\phi}\right)\big|_o}{\left(\alpha h_{\theta}h_{\phi}\right)\big|_{\varepsilon_1}}\right] \simeq 1
\end{equation}
which vanishes as well if the local scale is tiny compared to the gravitational radius. One could also define:
\begin{equation}
    \mathscr{C}=\frac{\left(\alpha h_{\theta}h_{\phi}\right)\big|_o}{\left(\alpha h_{\theta}h_{\phi}\right)\big|_i}\simeq 1
\end{equation}
When considering the limit of low magnetization ($\sigma_0\ll 1$, hence $\gamma_{\rm out}\simeq 1$ and $\gamma_{\rm in}\simeq 1$), the reconnection rate, under the effect of gravitation, obeys:
\begin{equation}
    R\equiv \frac{v_{\rm in}}{v_{\rm A}}\simeq \mathscr{CD}\frac{\delta}{L}\simeq \mathscr{C}^{-1}\frac{\eta}{\delta v_{\rm A}}\simeq \mathscr{D}^{1/2}S^{-1/2}
\end{equation}
\end{itemize}

From the results in this section, one can figure out that the properties of magnetic reconnection in Sweet-Parker model would be modified by gravitational effect but the modification is infinitesimal which disappears when $\delta\ll L\ll 1$, namely when the local scale is tiny compared to the gravitational radius. While the modifications argued in Ref.~\cite{CA2017}, which are still finite even though a tiny local scale is assumed, are actually generated from the substitutions of derivatives by finite differences in a wrong way.

\subsection{Plasma laboratory on equatorial plane}
\label{sec:plasma}

One can also consider the magnetic reconnection happening in the fluid's rest frame moving (call plasma laboratory henceforth) in the same way. As mentioned in Sect.~\ref{sec:intro}, magnetic reconnection is believed to occur in the plasma (described by ideal fluid in MHD scheme) moving around the compact object \cite{Yuan2024-1,Yuan2024-2}. Quantifying the process in the fluid's rest frame is of vital importance in the scenarios of magnetized accretion system. Here let us consider the current sheet posited parallel or perpendicular to the moving direction of plasma while $\hat{y}$ axis is chosen to be perpendicular to the equatorial plane. That is to say:
\begin{equation}
    k^{\mu}=e_{[0]}^{\mu};~~\chi^{\mu}=e_{[3]}^{\mu},~~\tau=e_{[2]}^{\mu},~~\zeta=e_{[1]}^{\mu}
    \label{eq:tetrad_p_para}
\end{equation}
for the current sheet parallel to the moving direction of plasma (para case), and:
\begin{equation}
    k^{\mu}=e_{[0]}^{\mu};~~\chi^{\mu}=e_{[1]}^{\mu},~~\tau=e_{[2]}^{\mu},~~\zeta=e_{[3]}^{\mu}
    \label{eq:tetrad_p_perp}
\end{equation}
for the current sheet perpendicular to the moving direction of plasma (perp case). Lee derivatives along $k^{\mu}$, $\chi^{\mu}$ and $\tau^{\mu}$ should be:
\begin{equation}
    \mathcal{L}_{\hat{k}}\approx \frac{\hat{\gamma}_s}{\alpha}\partial_t+\frac{\hat{\gamma}_s\hat{v}_s^{(r)}}{h_r}\partial_r+
    \hat{\gamma}_s\left(\frac{\hat{v}_s^{(\phi)}}{h_{\phi}}+\frac{\omega^{\phi}}{\alpha}\right)\partial_{\phi}
    \label{eq:steady_p}
\end{equation}
and
\begin{equation}
    \mathcal{L}_{\hat{\chi}}\approx
    \begin{cases}
        ~\frac{\hat{\gamma}_s\hat{v}_s}{\alpha}\partial_t+\frac{\hat{\gamma}_s\hat{v}_s^{(r)}}{\hat{v}_sh_r}\partial_r+\left(\frac{\hat{\gamma}_s\hat{v}_s^{(\phi)}}{\hat{v}_sh_{\phi}}+\frac{\hat{\gamma}_s\hat{v}_s\omega^{\phi}}{\alpha}\right)\partial_{\phi}~,~&\text{para case}\\
        ~\frac{\hat{v}_s^{(\phi)}}{\hat{v}_s h_r}\partial_r-\frac{\hat{v}_s^{(r)}}{\hat{v}_s h_{\phi}}\partial_{\phi}~, ~~~~~~~~~~~~~~~~~~~~~~~~~~~~~~&\text{perp case}
    \end{cases}
    ~~~~,~~~~
    \mathcal{L}_{\hat{\tau}}\approx \frac{1}{h_{\theta}}\partial_{\theta}
    \label{eq:L_delta_p}
\end{equation}
respectively. One can also choose a current sheet posited arbitrarily on the equatorial plane like
\begin{equation}
    \chi^{\mu}=\cos\xi_B e_{[3]}^{\mu}+\sin\xi_B e_{[1]}^{\mu},~~~\zeta^{\mu}=\cos\xi_B e_{[1]}^{\mu}-\sin\xi_B e_{[3]}^{\mu}
\end{equation}
where $\xi_B$ is, called orientation angle \cite{CA2021}, the angle between $\chi^{\mu}$ and the moving direction $e_{[3]}^{\mu}$. The Lee derivative along $\chi^{\mu}$ should now be:
\begin{equation}
    \mathcal{L}_{\hat{\chi}}\approx 
    \cos\xi_B\left[\frac{\hat{\gamma}_s\hat{v}_s}{\alpha}\partial_t+\frac{\hat{\gamma}_s\hat{v}_s^{(r)}}{\hat{v}_sh_r}\partial_r+\left(\frac{\hat{\gamma}_s\hat{v}_s^{(\phi)}}{\hat{v}_sh_{\phi}}+\frac{\hat{\gamma}_s\hat{v}_s\omega^{\phi}}{\alpha}\right)\partial_{\phi}\right]+
    \sin\xi_B\left(\frac{\hat{v}_s^{(\phi)}}{\hat{v}_s h_r}\partial_r-\frac{\hat{v}_s^{(r)}}{\hat{v}_s h_{\phi}}\partial_{\phi}\right)
    \label{eq:L_delta_p_xi}
\end{equation}

Similar conclusions could be got if one applies Eq.~\eqref{eq:tetrad_p_para}--\eqref{eq:L_delta_p_xi} into Eq.~\eqref{eq:step1}--\eqref{eq:step7}, after some tedious works on reorganizing the resultant expressions. Let us have a check by exhibiting some examples.

For the para case of plasma laboratory, Eq.~\eqref{eq:step2} gives: 
\begin{equation}
    \begin{split}
        &\partial_t\left(\hat{\gamma}_s\hat{v}_sh_rh_{\theta}h_{\phi}B_x\right)+\partial_r\left(\frac{\hat{\gamma}_s\hat{v}_s^{(r)}}{\hat{v}_s}\alpha h_{\theta}h_{\phi}B_x\right)+\partial_{\theta}\left(\alpha h_rh_{\phi}B_y\right) \\
        &~~~~~~~~~~~~~~~~~~~~~~~+\partial_{\phi}\left[\hat{\gamma}_s\left(\hat{v}_sh_rh_{\theta}h_{\phi}\omega^{\phi}+\frac{\hat{v}_s^{(\phi)}}{\hat{v}_s}\alpha h_rh_{\phi}\right)B_x\right]=0 \\
        ~~~~~~~~~~~~~~~~~~~~~~~~~\\
        \xrightarrow{\div \alpha h_rh_{\theta}h_{\phi}}~~&\frac{1}{h_{\theta}}\partial_{\theta}B_y+\left[\frac{\hat{\gamma}_s\hat{v}_s}{\alpha}\partial_t+\frac{\hat{\gamma}_s\hat{v}_s^{(r)}}{\hat{v}_sh_r}\partial_r+\left(\frac{\hat{\gamma}_s\hat{v}_s^{(\phi)}}{\hat{v}_sh_{\phi}}+\frac{\hat{\gamma}_s\hat{v}_s\omega^{\phi}}{\alpha}\right)\partial_{\phi}\right]B_x\simeq 0
    \end{split}
    \label{eq:para_step2}
\end{equation}
While for the perp case, Eq.~\eqref{eq:step2} gives:
\begin{equation}
    \begin{split}
        &\partial_r\left(\frac{\hat{v}_s^{(\phi)}}{\hat{v}_s}\alpha h_{\theta}h_{\phi}B_x\right)+\partial_{\theta}\left(\alpha h_rh_{\phi}B_y\right)-\partial_{\phi}\left(\frac{\hat{v}_s^{(r)}}{\hat{v}_s}\alpha h_rh_{\theta}B_x\right)=0 \\
        \xrightarrow{\div \alpha h_rh_{\theta}h_{\phi}}~~&\frac{1}{h_{\theta}}\partial_{\theta}B_y+\left(\frac{\hat{v}_s^{(\phi)}}{\hat{v}_s h_r}\partial_r-\frac{\hat{v}_s^{(r)}}{\hat{v}_s h_{\phi}}\partial_{\phi}\right)B_x \simeq 0
    \end{split}
    \label{eq:perp_step2}
\end{equation}
Both of the resultant equations above conclude $b_0/L\simeq b_1/\delta$ under the approximation argued in Eq.~\eqref{eq:temp} after utilizing the substitutions in Eq.~\eqref{eq:L_delta_GR}. 

Let us briefly summarize the results in Sect.~\ref{sec:zamo} and \ref{sec:plasma} here. From specific examples, one could figure out that equations in Eq.~\eqref{eq:step1}--\eqref{eq:step7} would provide similar conclusions as concluded from the SR versions introduced in Eq.~\eqref{eq:s1}--\eqref{eq:s7} if the quasi-stationary condition in Eq.~\eqref{eq:steady_GR} and the substitutions in Eq.~\eqref{eq:L_delta_GR} are chosen. Modifications of properties coming from gravitational effects vanish when $\delta\ll L\ll 1$. As already mentioned in Sect.~\ref{sec:setup}, basic equations of MHD in GR have no difference to their SR version when the first order derivatives of $g_{\mu\nu}$, physically speaking the tidal effect, are neglected. It is weird to say the spacetime curvature induces modifications on the properties in Sweet-Parker model since no higher order derivative of $g_{\mu\nu}$ and no non-linear term of affine connection are involved in the scheme of MHD, which could be checked from Eq.~\eqref{eq:GR-MHD2}. Here, I would like to reiterate that conclusions in Ref.~\cite{CA2017}, mentioning the properties in Sweet-Parker model would be modified by spacetime curvature effect, were actually made by utilizing incorrect substitutions like what Eq.~\eqref{eq:CA} showed.

\subsection{Plasma laboratory observed by ZAMOs}
\label{sec:p_zamo}

In the following two sections, let us have a look at some specific examples about the effect from observation. As introduced in Sect.~\ref{sec:obs}, although the gravitation takes almost no effect, relations in Sweet-Parker model would be modified if they are detected by an observer which is not at rest with respect to the laboratory. 

Considering in this section that the magnetic reconnection occurs in the plasma laboratory and is observed by ZAMOs on the equatorial plane. Namely we choose:
\begin{equation}
    k^{\mu}=e_{[0]}^{\mu}~~~,~~~n^{\mu}=\hat{e}_{(t)}^{\mu}
\end{equation}
Now one just needs to repeat the calculations in Eq.~\eqref{eq:v_obs}--\eqref{eq:rela_obs}. 

For the para case, in which $\chi^{\mu}=e_{[3]}^{\mu}$, $\tau^{\mu}=e_{[2]}^{\mu}$ and $\zeta^{\mu}=e_{[1]}^{\mu}$, Eq.~\eqref{eq:v_obs}--\eqref{eq:rela_obs} result in:
\begin{itemize}
\item {\bf Observed inflow and outflow speeds}

Eq.~\eqref{eq:v_obs} gives:
\begin{equation}
    \begin{split}
        {\rm inflow:}&~~v_{\rm in,obs}=\left|\frac{v_{\rm in}}{\hat{\gamma}_s}\hat{e}_{(\theta)}^{\mu}\right|=\hat{\gamma}_s^{-1}v_{\rm in} \\
        {\rm outflow:}&~~v_{\rm out,obs}=\left|\frac{v_{\rm out}\left(1-\hat{v}_s^2\right)}{1+v_{\rm out}\hat{v}_s}\left(\frac{\hat{v}_s^{(r)}}{\hat{v}_s}\hat{e}_{(r)}^{\mu}+\frac{\hat{v}_s^{(\phi)}}{\hat{v}_s}\hat{e}_{(\phi)}^{\mu}\right)\right|\simeq 
        \begin{cases}
            ~\left(1-\hat{v}_s^2\right)v_{\rm out}~~~&\text{LM} \\
            ~\left(1-\hat{v}_s\right)v_{\rm out}~~~&\text{HM}
        \end{cases}
    \end{split}
    \label{eq:v_obs1}
\end{equation}

\item {\bf Observed length of current sheet}

Eq.~\eqref{eq:L_obs} gives:
\begin{equation}
    L_{\rm obs}=\frac{L}{\left|\frac{\hat{\gamma}_s\hat{v}_s^{(r)}}{\hat{v}_s}\hat{e}_{(r)}^{\mu}+\frac{\hat{\gamma}_s\hat{v}_s^{(\phi)}}{\hat{v}_s}\hat{e}_{(\phi)}^{\mu}\right|}=\hat{\gamma}_s^{-1}L
    \label{eq:L_obs1}
\end{equation}
which is just the widely known length contraction in SR.

\item {\bf Observed Alfv{\'e}n velocity}

Eq.~\eqref{eq:vA_obs} gives:
\begin{equation}
    \sigma_{\rm 0,obs}=\frac{1}{\omega}\left(\frac{\hat{v}_s^{(r)}b_0}{\hat{v}_s}\hat{e}_{(r)}^{\mu}+\hat{\gamma}_s\hat{v}_se_z\hat{e}_{(\theta)}^{\mu}+\frac{\hat{v}_s^{(\phi)}b_0}{\hat{v}_s}\hat{e}_{(\phi)}^{\mu}\right)^2\simeq \sigma_0~~\rightarrow~~v_{\rm A,obs}\simeq v_{\rm A}
    \label{vA_obs1}
\end{equation}
where we adopt $v_{\rm in}\ll 1$ such that $e_z\ll b_0$.

\item {\bf Observed relations}

Conclusively, Eq.~\eqref{eq:rela_obs} gives:
\begin{equation}
    \frac{R_{\rm obs}}{S_{\rm obs}^{1/2}}\simeq \hat{\gamma}_s^{-3/2}\frac{R}{S^{-1/2}}~~~,~~~
    \frac{v_{\rm out,obs}}{v_{\rm A,obs}}\simeq 
    \begin{cases}
        ~\left(1-\hat{v}_s^2\right)\frac{v_{\rm out}}{v_{\rm A}}~~~&\text{LM} \\
        ~\left(1-\hat{v}_s\right)\frac{v_{\rm out}}{v_{\rm A}}~~~~&\text{HM}
    \end{cases}
\end{equation}
by utilizing the results in Eq.~\eqref{eq:v_obs1}, \eqref{eq:L_obs1} and \eqref{vA_obs1}.
\end{itemize}

From the calculations above, it could be concluded that when a slow magnetic reconnection which obeys the rules in Sweet-Parker model occurs in the rest frame of plasma moving on equatorial plane with current sheet posited along the moving direction and is detected by ZAMOs, the observer may find out $R/S^{-1/2}\simeq \hat{\gamma}_s^{-3/2}$ instead of $R/S^{-1/2}\simeq 1$ and $v_{\rm out}/v_{\rm A}\simeq 1-\hat{v}_s^2$ (for LM) or $v_{\rm out}/v_{\rm A}\simeq 1-\hat{v}_s$ (for HM) instead of $v_{\rm out}/v_{\rm A}\simeq 1$ from detections.

For the general case, in which $\chi^{\mu}=\cos\xi_B e_{[3]}^{\mu}+\sin\xi_B e_{[1]}^{\mu}$, $\tau^{\mu}=e_{[2]}^{\mu}$ and $\zeta^{\mu}=\cos\xi_B e_{[1]}^{\mu}-\sin\xi_B e_{[3]}^{\mu}$, Eq.~\eqref{eq:v_obs}--\eqref{eq:rela_obs} result in:
\begin{itemize}
\item {\bf Observed inflow and outflow speed}

Eq.~\eqref{eq:v_obs} gives:
\begin{equation}
    \begin{split}
        {\rm inflow:}&~~v_{\rm in,obs}=\left|\frac{v_{\rm in}}{\hat{\gamma}_s}\hat{e}_{(\theta)}^{\mu}\right|=\hat{\gamma}_s^{-1} v_{\rm in} \\
        {\rm outflow:}&~~v_{\rm out,obs}=...= \frac{\left(\hat{\gamma}_s^{-2}\cos^{2}\xi_B+\sin^2\xi_B\right)^{1/2}}{\hat{\gamma}_s\left(1+v_{\rm out}\hat{v}_s\cos\xi_B\right)}v_{\rm out}\simeq
        \begin{cases} 
            ~\hat{\gamma}_s^{-2}\left(\cos^{2}\xi_B+\hat{\gamma}_s^2\sin^2\xi_B\right)^{1/2}~~~&\text{LM} \\
            ~\frac{\left(\cos^{2}\xi_B+\hat{\gamma}_s^2\sin^2\xi_B\right)^{1/2}}{\hat{\gamma}_s^2\left(1+\hat{v}_s\cos\xi_B\right)}v_{\rm out}~~~~~~~~~&\text{HM}
        \end{cases}
    \end{split}
\end{equation}

\item {\bf Observed length of current sheet}

Eq.~\eqref{eq:L_obs} gives:
\begin{equation}
    L_{\rm obs}=\frac{L}{\left|\frac{\hat{\gamma}_s\hat{v}_s^{(r)}\cos\xi_B+\hat{v}_s^{(\phi)}\sin\xi_B}{\hat{v}_s}\hat{e}_{(r)}^{\mu}+\frac{\hat{\gamma}_s\hat{v}_s^{(\phi)}\cos\xi_B-\hat{v}_s^{(r)}\sin\xi_B}{\hat{v}_s}\hat{e}_{(\phi)}^{\mu}\right|}=\frac{L}{\left(\hat{\gamma}_s^2\cos^2\xi_B+\sin^2\xi_B\right)^{1/2}}
\end{equation}

\item {\bf Observed Alfv{\'e}n velocity}

Eq.~\eqref{eq:vA_obs} gives:
\begin{equation}
    \sigma_{\rm 0,obs}=...\simeq \left(\cos^2\xi_B+\hat{\gamma}_s^2\sin^2\xi_B\right)\sigma_0~~\rightarrow~~
    v_{\rm A,obs}\simeq 
    \begin{cases}
        ~\left(\cos^2\xi_B+\hat{\gamma}_s^2\sin^2\xi_B\right)^{1/2}v_{\rm A}~~~&\text{LM} \\
        ~v_{\rm A}~~~~~~~~~~~~~~~~~~~~~~~~~~~~~~~~~~~&\text{HM}
    \end{cases}
\end{equation}
where we adopt $v_{\rm in}\ll 1$.

\item {\bf Observed relations}

Conclusively, Eq.~\eqref{eq:rela_obs} gives:
\begin{equation}
    \begin{split}
        \frac{R_{\rm obs}}{S_{\rm obs}^{1/2}}~\simeq~& \frac{R}{S^{-1/2}}\times
        \begin{cases}
            ~\hat{\gamma}_s^{-1}\left[\hat{\gamma}_s^2\left(\cos^4\xi_B+\sin^4\xi_B\right)+\left(1+\hat{\gamma}_s^4\right)\cos^2\xi_B\sin^2\xi_B\right]^{-1/4}~~~&\text{LM} \\
            ~\hat{\gamma}_s^{-1}\left(\hat{\gamma}_s^2\cos^2\xi_B+\sin^2\xi_B\right)^{-1/4}~~~~~~~~~~~~~~~~~~~~~~~~~~~~~~~~~~~~~~~&\text{HM}
        \end{cases} \\
        ~~~~~~~~~~&~~~~~~~~~~~~~~~~~~~\\
        \frac{v_{\rm out,obs}}{v_{\rm A,obs}}~\simeq~& \frac{v_{\rm out}}{v_{\rm A}}\times
        \begin{cases}
            ~\hat{\gamma}_s^{-2}~~~~~~~~~~~~~~~~~~~~~~~~&\text{LM}  \\
            ~\frac{\left(\cos^{2}\xi_B+\hat{\gamma}_s^2\sin^2\xi_B\right)^{1/2}}{\hat{\gamma}_s^2\left(1+\hat{v}_s\cos\xi_B\right)}~~~&\text{HM}
        \end{cases}
    \end{split}
\end{equation}
\end{itemize}

It is easily seen that the results are the same as those in para case when $\xi_B=0$. One can also check that the results are the same as those in perp case if $\xi_B=\pi/2$.

\subsection{ZAMOs laboratory observed by static observer}
\label{sec:zamo_t}

Let us now consider the ZAMOs laboratory on the equatorial plane with current sheet posited azimuthally, as what Sect.~\ref{sec:zamo} described. The observer is set to be static in BL coordinates. Namely, we choose:
\begin{equation}
    k^{\mu}=\hat{e}_{(t)}^{\mu}~~~,~~~n^{\mu}=\left[\alpha^2-\left(h_{\phi}\omega^{\phi}\right)^2\right]^{-1/2}\partial_t^{\mu}
\end{equation}
Notice that this observer is well defined outside the ergosphere only. Now let us repeat the calculations in Eq.~\eqref{eq:v_obs}--\eqref{eq:rela_obs} again:
\begin{itemize}
\item {\bf Observed inflow and outflow speeds}

Eq.~\eqref{eq:v_obs} gives:
\begin{equation}
    \begin{split}
        {\rm inflow:}&~~v_{\rm in,obs}=...=\alpha^{-1}\left[\alpha^2-\left(h_{\phi}\omega^{\phi}\right)^2\right]^{1/2}v_{\rm in} \\
        {\rm outflow:}&~~v_{\rm out,obs}=...=\frac{\alpha^2-\left(h_{\phi}\omega^{\phi}\right)^2}{\alpha\left(\alpha+h_{\phi}\omega^{\phi}v_{\rm out}\right)^{1/2}}v_{\rm out}\simeq
        \begin{cases}
            ~\left[1-\alpha^{-2}\left(h_{\phi}\omega^{\phi}\right)^2\right]v_{\rm out}~~~&\text{LM} \\
            ~\left(1-\alpha^{-1}h_{\phi}\omega^{\phi}\right)v_{\rm out}~~~~~~~&\text{HM}
        \end{cases}
    \end{split}
\end{equation}

\item {\bf Observed length of current sheet}

Eq.~\eqref{eq:L_obs} gives:
\begin{equation}
    L_{\rm obs}=...=\alpha^{-1}\left[\alpha^2-\left(h_{\phi}\omega^{\phi}\right)^2\right]^{1/2}L
\end{equation}

\item {\bf Observed Alfv{\'e}n velocity}

Eq.~\eqref{eq:vA_obs} gives:
\begin{equation}
    \sigma_{\rm 0,obs}=...\simeq \frac{\alpha^2-\left(h_{\phi}\omega^{\phi}\right)^2}{\alpha^2}\sigma_0~~\rightarrow~~
    v_{\rm A,obs}\simeq 
    \begin{cases}
        ~\alpha^{-1}\left[\alpha^2-\left(h_{\phi}\omega^{\phi}\right)^2\right]^{1/2}v_{\rm A}~~~&\text{LM} \\
        ~v_{\rm A}~~~~~~~~~~~~~~~~~~~~~~~~~~~~~~~~~&\text{HM}
    \end{cases}
\end{equation}

\item {\bf Observed relations}

Conclusively, Eq.~\eqref{eq:rela_obs} gives:
\begin{equation}
    \begin{split}
        \frac{R_{\rm obs}}{S_{\rm obs}^{1/2}}~\simeq~& \frac{R}{S^{-1/2}}\times
        \begin{cases}
            ~\alpha^{-1}\left[\alpha^2-\left(h_{\phi}\omega^{\phi}\right)^2\right]^{1/2}\simeq 1-2a^2r^{-4}-4a^2r^{-5}+...~~~&\text{LM} \\
            ~\alpha^{-3/2}\left[\alpha^2-\left(h_{\phi}\omega^{\phi}\right)^2\right]^{3/4}\simeq 1-3a^2r^{-4}-6a^2r^{-5}+...~~~&\text{HM}
        \end{cases} \\
        ~~~~~~~~~~&~~~~~~~~~~~~~~~~~~~\\
        \frac{v_{\rm out,obs}}{v_{\rm A,obs}}~\simeq~& \frac{v_{\rm out}}{v_{\rm A}}\times
        \begin{cases}
            ~\alpha^{-1}\left[\alpha^2-\left(h_{\phi}\omega^{\phi}\right)^2\right]^{1/2}\simeq 1-2a^2r^{-4}-4a^2r^{-5}+...~~~&\text{LM}  \\
            ~1-\alpha^{-1}h_{\phi}\omega^{\phi}\simeq 1-2ar^{-2}-2ar^{-3}+...~~~~~~~~~~~~~~~~~~&\text{HM}
        \end{cases}
    \end{split}
    \label{eq:rela_z_1}
\end{equation}
\end{itemize}

From Eq.~\eqref{eq:rela_z_1} one could find out that the detected ratios decrease for both the reconnection rate and outflow speed, except for the outflow speed in the high magnetization limit with positive $a$, when an static observer gets closer and closer to the BH. Recall that the observer is well defined outside the ergo sphere only. In this sense, one always has $r>2$ on the equatorial plane. As a consequence, the observer could neither get a negative reconnection rate nor a negative outflow speed. That is to say, no counterintuitive value would be got from detection.

\section{Summary}
\label{sec:sum}

In this work, I am trying to explore an answer about how to describe the Sweet-Parker model, one of the most famous and widely used theoretical models of magnetic reconnection, in general relativity. 

I start from recapping the calculations in Sweet-Parker model, which are reorganized into seven steps, briefly. And then I make some basic setups for preparations, including the basic equations of MHD in GR, the rest frames in which the magnetic reconnection occurs or is observed, the spacelike vectors for determining how the current sheet is posited and structure of electromagnetic field in the local scale. Most importantly, I propose the GR form of quasi-stationary condition and the approximations which substitute derivatives in equations by finite differences over the local scale. With those preparations, I subsequently propose the GR forms of the seven steps in Sweet-Parker model and discuss the probable modifications of relations in Sweet-Parker model induced from observation in general. I exhibit the calculations for specific laboratories and observers, showing the probable infinitesimal modifications coming from gravitation effect and the probable nontrivial modifications caused by observation. 

I compare my results to the results in Ref.~\cite{CA2017} and reiterate my insistence that spacetime curvature would never provide significant modification on the properties of magnetic reconnection and the modifications caused by gravitational effect should vanish when the tidal effect is neglected for the local scale is tiny (compared to the gravitational radius), which is opposite to the conclusions in Ref.~\cite{CA2017}. While the relations in Sweet-Parker model would be inevitably modified significantly if the magnetic reconnection process is detected by an observer which is not at rest with respect to the laboratory in the rest frame of which the magnetic reconnection occurs.

\section*{Acknowledgement}

I appreciate valuable discussions and constructive suggestions given by Prof.~Bin Chen. I would like to thank Dr.~Yuedan Wang for her encouragement.

\appendix

\newpage
\bibliographystyle{utphys}
\bibliography{references}

\end{document}